\documentclass[twocolumn]{aastex62}

\def\ltsima{$\; \buildrel < \over \sim\;$}
\def\ltsim{\lower.5ex\hbox{\ltsima}}
\def\gtsima{$\; \buildrel > \over\sim \;$}
\def\gtsim{\lower.5ex\hbox{\gtsima}}
\def\ms{$M_{\odot}$ }
\def\msp{$M_{\odot}$}

\received{xxx}
\revised{xxx}
\accepted{xxx}

\shorttitle{r-process enrichment in the Galactic halo}
\shortauthors{Tsujimoto, Nishimura, \& Kyutoku}

\begin{document}

\title{r-Process enrichment in the Galactic halo characterized by nucleosynthesis variation \\ in the ejecta of coalescing neutron star binaries}

\correspondingauthor{Takuji Tsujimoto}
\email{taku.tsujimoto@nao.ac.jp}

\author[0000-0002-9397-3658]{Takuji Tsujimoto}
\affil{National Astronomical Observatory of Japan, Mitaka, Tokyo 181-8588, Japan}

\author[0000-0002-0842-7856]{Nobuya Nishimura}
\affiliation{Center for Gravitational Physics, Yukawa Institute for Theoretical Physics, Kyoto University, Kyoto 606-8502, Japan}

\author[0000-0003-3179-5216]{Koutarou Kyutoku}
\affiliation{Department of Physics, Kyoto University, Kyoto 606-8502, Japan}
\affiliation{Center for Gravitational Physics, Yukawa Institute for Theoretical Physics, Kyoto University, Kyoto 606-8502, Japan}
\affiliation{Theory Center, Institute of Particle and Nuclear Studies, KEK, Tsukuba 305-0801, Japan}
\affiliation{Interdisciplinary Theoretical and Mathematical Sciences Program (iTHEMS), RIKEN, Wako, Saitama 351-0198, Japan}



\begin{abstract}
A large star-to-star variation in the abundances of r-process elements, as seen in the [Eu/Fe] ratio for Galactic halo stars, is a prominent feature that is distinguishable from other heavy elements. It is, in part, caused by the presence of highly r-process enriched stars, classified as r-II stars ([Eu/Fe]$\geq+1$). In parallel, halo stars show that the ratio of a light r-process element (Y) to Eu is tightly correlated with [Eu/Fe], giving the lowest [Y/Eu] ratio that levels off at r-II stars. On the other hand, recent hydrodynamical simulations of coalescing double neutron stars (cNSNSs) have suggested that r-process sites may be separated into two classes providing different electron-fraction distributions: tidally-driven dynamical ejecta and (dynamical or postmerger) non-tidal ejecta. Here, we show that a widely spanning feature of [Eu/Fe] can be reproduced by models that consider the different masses of tidally-driven dynamical ejecta from both cNSNSs and coalescing black hole/neutron star binaries (cBHNSs). In addition, the observed [Y/Eu] trend is explained by the combined nucleosynthesis in two kinds of ejecta with varying mass asymmetry in double NS systems. Our scenario suggests that massive tidally-driven dynamical ejecta accompanied by massive non-tidal part from cNSNSs or cBHNSs could alone accommodate r-II abundances, including an actinide boost in some cases. The event rate for cNSNSs estimated from our study agrees with the latest result of $\sim1000$ (90\% confidence interval of $110-3840$) Gpc$^{-3}$yr$^{-1}$ by gravitational-wave detection, and a few events per Gpc$^{3}$ per year of cBHNSs associated with r-process production are predicted to emerge.

\end{abstract}



\section{Introduction}

Clear answers to the origin and evolution of r-process elements continue to elude us, even after the discovery of gravitational waves from coalescing double neutron star (cNSNS) GW170817 and the subsequent discovery of multi-wavelength electromagnetic counterparts that identified cNSNSs as a promising major source of r-process elements \citep[e.g.,][]{Coulter_17, Smartt_17, Pian_17, Cowperthwaite_17, Thielemann_17}. The stellar record of r-process abundances leaves open the question of whether cNSNSs are the sole (major) site of the r-process. This is exemplified by two arguments: that the r-process abundance features of very metal-poor stars invoke the contribution from some specific core-collapse supernovae (CCSNe), the so-called magneto-rotational SNe \citep[e.g.,][]{Winteler_12, Wehmeyer_15, Nishimura_15}, and that disk stars suggest enrichment by another rare CCSNe, collapsars, as the major site of the r-process \citep{MacFadyen_99, Fujimoto_08, Siegel_19}. However, it is not clear whether these specific CCSNe exist if we consider the propagation of strong-magnetic jets in the models for magneto-rotational SNe \citep{Nishimura_17, Halevi_18, Mosta_18} and constraints from stellar abundances of halo stars on collapsars \citep{Macias_19}.

A large scatter seen in the [Eu/Fe] ratio, spanning nearly three orders of magnitude among Galactic halo stars, is a feature unique to r-process elements and thus should be connected to r-process sites and their nucleosynthesis \citep[e.g.,][]{Cowan_19}; however, the origin of the scatter has yet to be identified. Among Eu-measured halo stars, special attention has been directed to those stars having highly enhanced Eu abundances, as high as [Eu/Fe]$> +1$. These have been  classified as r-II stars \citep{Beers_05} since the discovery of the first r-II star, CS 22892-052 \citep{Sneden_94}. Recent first detection of r-II stars outside the Galaxy in an ultra-faint dwarf galaxy Reticulumn II \citep[Ret II,][]{Ji_16} triggered one possible scenario for the origin of Galactic r-II stars; i.e., these stars are accreted from disrupted small satellite galaxies \citep{Roederer_18, Brauer_19}. However, as seen in their orbital property that about 70 \% of r-II stars reside within the inner regions ($<13$ kpc) of the Galactic halo \citep{Roederer_18}, we have no clear kinematic evidence for their accretion history. Alternatively, observations suggest the possibility that r-II stars differ from other stars in the nucleosynthesis condition of the associated r-process events. These r-II stars are likely to be born in an environment that is neutron-rich enough to produce an actinide (e.g., Th and U) boost for about one third of them \citep{Mashonkina_14}, although the conditions for the actinide production involve uncertainties, including the possibility that actinides could be populated at the not that low electron fraction ($Y_e$) \citep[e.g.,][]{Lippuner_17}.

There is another implication for the connection between the variation in [Eu/Fe] and the nucleosynthesis of r-process events. The abundance ratios of light r-process elements (Sr, Y, and Zr)\footnote{Although the s-process dominates these neutron-capture elements in the solar abundance pattern, we focus on the chemical enrichment in the Galactic halo where the r-process is the major source for these elements in this study. Therefore, light neutron-capture elements are hereafter referred to as light r-process elements.} to Eu are correlated with [Eu/Fe] in the sense that [light r-process/Eu] is larger for smaller [Eu/Fe] \citep{Montes_07, Tsujimoto_14a}. We need to answer why r-II stars show the lowest [light r-process/Eu] ratio among halo stars.

An astrophysical site that is different for light r-process elements than for heavy ones including Eu, has been proposed as the mechanism behind the variation in [light r-process/Eu] for individual stars \citep{Montes_07, Francois_07}.  Possible candidates include regular CCSNe producing Fe together \citep{Qian_07} through the neutrino-driven wind from the proto-NS \citep{Arcones_14}, electron-capture SNe \citep{Kitaura_06, Wanajo_11}, and magneto-rotational SNe \citep{Nishimura_17}. However, our understanding to date should be reexamined in response to recent numerical simulations for r-process nucleosynthesis in cNSNSs. The latest discovery of a signature of synthesized  light r-process element (Sr) in the afterglow of GW170817 strengthens this necessity \citep{Watson_19}.

Currently, binary NSs are regarded as being capable of producing various kinds of ejecta during the coalescence. Tidal interaction could allow the NS material to be ejected without experiencing significant shock heating nor neutrino irradiation \citep[e.g.,][]{Freiburghaus_99, Sekiguchi_16, Vincent_19}, although recent simulations have reported a non-negligible influence of neutrinos on such ejecta \citep{Radice_18}. These tidal ejecta may keep low $Y_e$ and contribute predominantly to the production of heavy r-process elements such as Eu. Dynamical interaction also produces the shock-heated ejecta, whose $Y_e$ is suggested by some studies to be increased to the extent that Eu is no longer produced efficiently and instead light r-process elements such as Y are produced \citep[e.g.,][]{Wanajo_14}. Following these dynamical ejecta, the merger remnant drives postmerger mass ejection via various possible mechanisms \citep[e.g.,][]{Metzger_14, Just_15, Siegel_18, Fujibayashi_18}. In particular, \citet{Fujibayashi_18} find that the postmerger wind from the torus surrounding a massive NS, which is likely to be formed for typical cNSNS, is primarily characterized by moderate $Y_e$ values and dominantly produces light r-process elements. These updated theoretical inputs of cNSNS may have high potential for altering the interpretation of [light r-process/Eu] variation. 

A close connection between the star-to-star variation in r-process abundances and cNSNSs is also implied from the wide range of [Eu/Fe] ratios. The mass of dynamical ejecta is predicted to vary as widely as two orders of magnitude \citep[$\sim10^{-4}-0.01$ \ms:][$\sim2\times10^{-4}-0.02$ \ms: Bauswein et al.~2013]{Hotokezaka_13}, where {\it the tidally-driven component} increases as the mass asymmetry increases. Further, its mass range widens if we consider the contribution from coalescing black hole/neutron star binaries (cBHNSs); i.e., the mass of dynamical ejecta from cBHNSs can become as massive as $\sim 0.1$ \ms dominated completely by the tidal component \citep{Kyutoku_15}. As a result, the ejected Eu mass can differ by approximately three orders of magnitude according to the difference in the mass of its production site. Then, it turns out that the degree of variation in an Eu mass broadly coincides with that for the observed [Eu/Fe] ratios. Indeed, in the Draco dwarf galaxy, we identified two r-process events with enriched gases at levels that differ by more than one order of magnitude \citep{Tsujimoto_17a}. These arguments suggest that enrichment driven by both cNSNSs and cBHNSs makes a fundamental feature of Galactic r-process abundance, including the early onset of the r-process at low metallicities \citep{Wehmeyer_19}.

\section{\bf Nucleosynthesis in tidal and non-tidal ejecta}

In this study, we boldly classify the ejecta components into the tidal and non-tidal ejecta, which are responsible for the synthesis of Eu and Y, respectively. Such a classification is possible since the production sites of Eu and Y inside the ejecta can be seen as discrete in terms of the $Y_{\rm e}$ value. To investigate the dependence of nucleosynthesis yields on $Y_{\rm e}$, we perform a set of r-process calculations, using a nuclear reaction network code \citep{Nishimura_16, Nishimura_17} and the simplified models of merger ejecta for cNSNSs and cBHNSs. We adopt the adiabatic free-expansion evolution \citep{Freiburghaus_99, Farouqi_10, Wanajo_18}, of which the abundance evolution is determined by the initial $Y_{\rm e}$ value, entropy, and expansion velocity.

The results of nucleosynthesis calculations are shown in Figure 1, in which the mass fractions of Eu and Y are plotted as a function of $Y_{\rm e}$. The mass fraction has a band between the minimum and maximum values for the varieties of  entropy: 10--35 $k_{\rm B}~{\rm baryon}^{-1}$ and expansion velocity: $0.05$--$0.3$ of the speed of light. These adopted values cover the reasonable range for merger conditions/ejecta suggested by \cite{Wanajo_18}. We see a clear separation of Eu and Y production depending on the value of $Y_e$. The deviation of the mass fractions of Eu and Y lies within a factor of $\sim 2$ for $Y_{\rm e} < 0.3$. It suggests that the final abundance of Eu, which is a key element of the tidal ejecta, is primarily  determined by $Y_{\rm e}$ while the Y abundance shows a bit more complexity of a dependence on other factors. In addition, the uncertainty of the nuclear physics input such as the mass model and fission treatment \citep{Mendoza_15, Goriely_15, Eichler_15, Cote_18, Vassh_19} affects calculated abundances of products such as Eu since dynamical ejecta conditions can synthesize nuclei well outside the reach of current experimental data.

\begin{figure}[t]
	\includegraphics[width=\columnwidth]{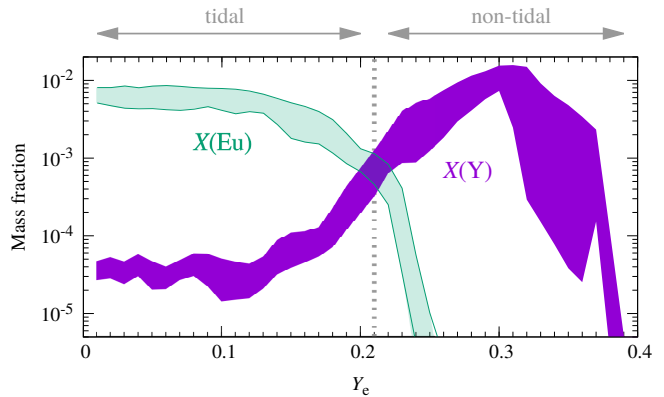}
\caption{The mass fractions of Eu and Y by the r-process as a function of $Y_{\rm e}$. The nucleosynthesis calculations \citep{Nishimura_17} are performed with the parametric expansion model of merger ejecta \citep{Wanajo_18}. The mass fraction has a band between the minimum and maximum values for the varieties of entropy: 10--35 $k_{\rm B}~{\rm baryon}^{-1}$ and expansion velocity: $0.05$--$0.3$ of the speed of light. $Y_{\rm e} \sim 0.22$ is the boundary of the tidal and non-tidal components assumed in this study.}
\end{figure}

We also need to keep in mind that the nucleosynthesis yields in both the dynamical and postmerger ejecta are inevitably dependent on many uncertain physics inputs in hydrodynamical simulations. Examples include neutrino transport scheme \citep{Caballero_12, Foucart_16, Radice_18}, neutrino oscillation \citep{Malkus_16, Frensel_17, Tian_17}, and neutron star equations of state \citep{Sekiguchi_15, Bovard_17, Radice_18, Vincent_19}. In particular, the r-process yields associated with the postmerger ejecta are critically important, because they could dominate the mass of ejecta in many cases, given the wide range of accretion disk mass ejection deduced \citep{Cote_18}. It is possible that the postmerger ejecta also efficiently produce Eu due to low $Y_e$ if the BH is formed promptly formed after merger \citep{Just_15, Siegel_18}.

\section{Variations in [E{\scriptsize u}/F{\scriptsize e}] and [Y/E{\scriptsize u}]}

We discuss more comprehensively the origin of variations in both [Eu/Fe] and [light r-process/Eu] and the correlation between the two. It was first found by \citet{Montes_07} that there exists a downward trend of [light r-process/Eu] with increasing [Eu/Fe] while [heavy r-process/Eu] stays constant with respect to [Eu/Fe] (here light r-process elements are Sr, Y and Zr and heavy ones include Ba, La, and Nd). This feature may suggest that the major source of light r-process elements is different from that of Eu \citep[see also][]{Aoki_05, Francois_07}, and is proposed to be explained by the presence of a weak r-process producing source such as the light element primary process (LEPP) as a consequence of the correlation of light r-process elements with Fe instead of Eu \citep{Travaglio_04, Montes_07}. 

\begin{figure}[t]
	\includegraphics[width=\columnwidth]{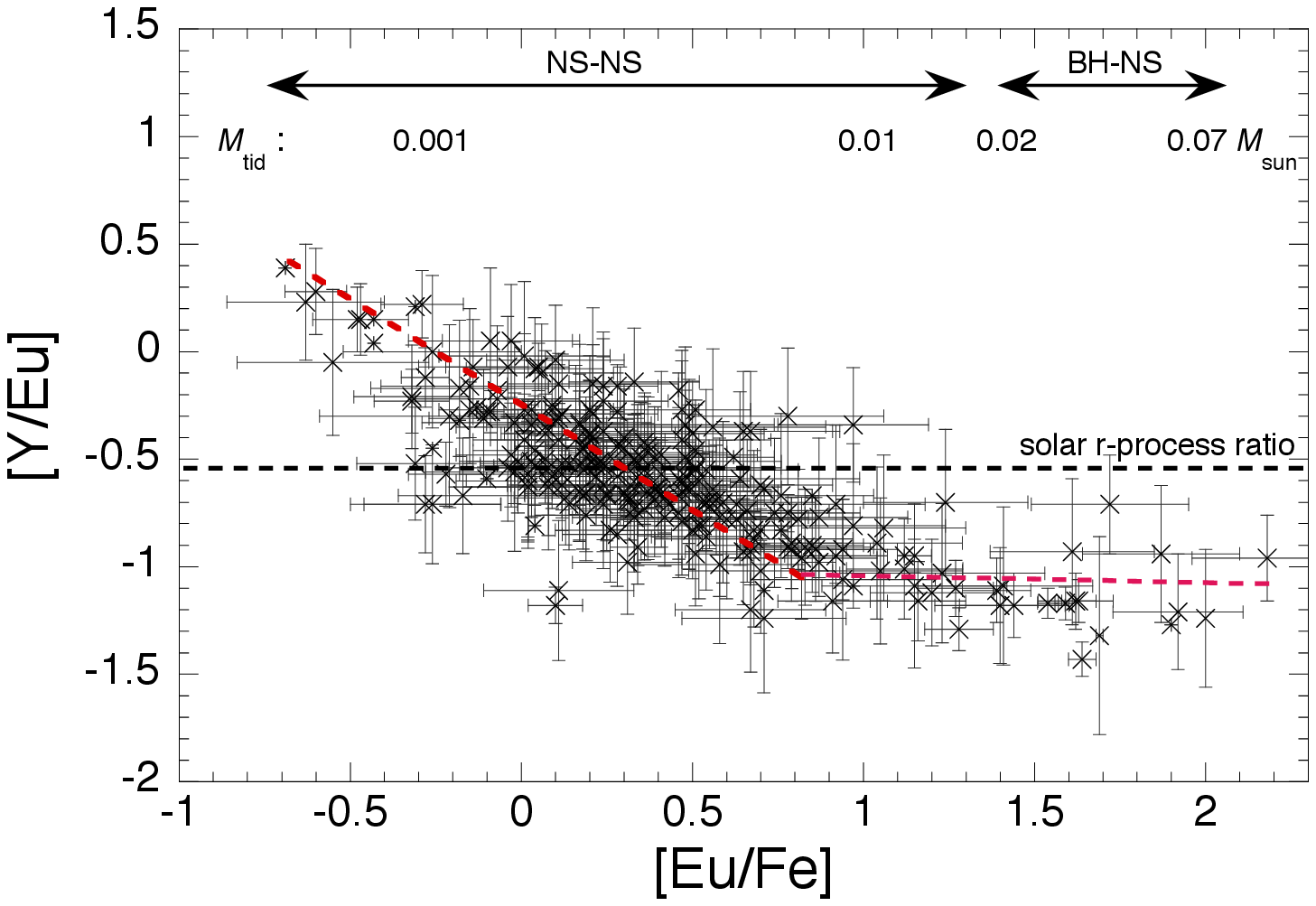}
\caption{Observed correlation of [Y/Eu] with [Eu/Fe] for Galactic halo stars that reflect the r-process Y/Eu abundance ratio. The observed data are selected with [Fe/H]$<-2$ and [Ba/Eu]$<$0 from a SAGA database \citep{Suda_08}. For r-II stars, the data are complemented by a JINA database \citep{Abohalima_18}. In addition, two r-II stars recently discovered \citep{Sakari_18, Holmbeck_18} are also added. One extremely Eu-enhanced star ([Eu/Fe]=+2.74 and [[Y/Eu]=$-0.84$; Allen et al.~2012) is not shown here. Black dashed line indicates the solar r-process ratio \citep{Simmerer_04, Bisterzo_14}. Brief explanations are attached in the upper part (see the text). Here the assumed higher mass ejecta of low $Y_{\rm e}$ tidal ejecta for cBHNS scenarios is motivated by numerical simulation results \citep{Kyutoku_15}. Red dashed line indicates a mean abundance track.}
\end{figure}

With an update of the observational data, we revisit this matter using Y for light r-process elements, since its abundance determination is generally more reliable than than that of Sr and Zr. To avoid the stars whose Y abundances include s-process contributions, we select halo stars that satisfy [Fe/H]$<-2$ and [Ba/Eu]$<0$. The [Y/Eu] vs.~[Eu/Fe] diagram thus obtained is shown in Figure 2. The overall correlation between [Y/Eu] and [Eu/Fe] is summarized as (i) a decreasing [Y/Eu] with increasing [Eu/Fe] for [Eu/Fe]\ltsim$+1$ and (ii) a broadly constant [Y/Eu] for [Eu/Fe]\gtsim$+1$. Thus, r-II stars give the lowest [Y/Eu] ($\approx-1.1$). 

In this section, we argue that this observed correlation may be naturally explained by assuming that variations in both [Eu/Fe] and [Y/Eu] for metal-poor halo stars reflect directly those in nucleosynthesis from the merger events. Specifically, the correlation between [Y/Eu] and [Eu/Fe] is regarded as the result of variations in the properties of cNSNSs or cBHNSs, for which Eu and Y are assumed to be synthesized separately in the tidal and non-tidal ejecta of individual merger events, respectively, as discussed in \S 2. Here, we assume the solar r-process pattern including the $A\sim 130$ peak (i.e., $Z\geq$52) with no production for $Z<$52 for the matter within the tidal ejecta from both cNSNSs and cBHNSs. On the other hand, for the non-tidal ejecta, we assume the nucleosynthesis pattern following the recent numerical results for the postmerger ejecta which show significantly low production of lanthanide elements and give the mass fraction of $1.64\times10^{-2}$ for Y \citep{Fujibayashi_18}.

We first propose that the degree of Eu enhancement indicated by [Eu/Fe] for individual stars ($-0.7$\ltsim[Eu/Fe] \ltsim+2) could be basically determined by the different levels of Eu enrichment of gas via associated merger events (cNSNSs or cBHNSs) with widely varying masses ($M_{\rm tid}$) of tidally-driven dynamical ejecta ($10^{-4}$\ms\ltsim $M_{\rm tid}$\ltsim$0.1$\msp). Considering the [Eu/Fe] distribution of halo stars, we identify two components: a Gaussian distribution in the range of $-0.7$\ltsim[Eu/Fe]\ltsim+1.3 with a peak around [Eu/Fe]$\approx+0.3$ and an extending Eu-rich tail up to [Eu/Fe]$\approx+2$. Then, considering the predicted mass range of dynamical ejecta, cNSNS and cBHNS are likely to be responsible for producing the former and latter distributions, respectively. Approximate mass values of the mass of tidally-driven dynamical ejecta are shown in the upper part of the figure, implying that the average value $\langle M_{\rm tid}\rangle$ for cNSNSs should be a few times $10^{-3}$ \msp. If cBHNS should produce more Eu than cNSNS due to larger tidal mass ejection, the Y/Eu plateau could be connected to such events. Considering very low production of light r-process elements in the tidally dominated dynamical ejecta of cBHNSs owing to a low $Y_e$ (\ltsim 0.1) \citep{Roberts_17}, we argue that Y in r-II stars originates from the postmerger ejecta. 

The Y/Eu plateau implies that the production ratio of Y to Eu is quite similar in individual cBHNS events. Since each element owes its origin to non-tidal postmerger ejecta and tidally-driven dynamical ejecta, respectively, the masses of the two type of ejecta are suggested to scale linearly with each other. This linearity may also hold even if the postmerger ejecta of cBHNSs produce Eu due to a smaller $Y_e$ resulting from lower neutrino irradiation \citep{Fernandez_17}, though there are complexities in the influence of the neutrino treatment \citep[e.g.,][]{Caballero_12, Wu_17}. The required condition of a similar mass ratio of light r-process elements to Eu among the postmerger ejecta of cBHNSs needs to be investigated \citep[see also][]{Kyutoku_15}.

On the other hand, the trend of an increasing [Y/Eu] with decreasing [Eu/Fe] at the low [Eu/Fe] regime is likely to reflect the diversity of nucleosynthesis in cNSNSs with various mass asymmetry in our scheme. The tidal interaction tends to be efficient for asymmetric binaries, and $M_\mathrm{tid}$ becomes large. At the same time, shock interactions are not very efficient for dynamical mass ejection in asymmetric binaries. Thus, the average values of $Y_e$ become lower for more asymmetric binaries \citep[e.g.,][]{Sekiguchi_16}. These combined effects could explain the observed Y/Eu trend. According to this view, cNSNSs with massive ($\gtrsim 0.01 M_\odot$) tidally-driven dynamical ejecta produce a very small amount of Y and share the Y/Eu plateau with cBHNSs. However, it is necessary to comprehensively investigate the dependence of nucleosynthesis in cNSNSs on the mass asymmetry and other binary parameters \citep{Kiuchi_19}, particularly in light of recently discovered highly asymmetric double NSs \citep{Ferdman_18}. In addition to the above discussion, a contrast in [Y/Eu] may, in part, be attributed to the dependence of Y nucleosynthesis in the postmerger ejecta on their different masses. As one possibility, we anticipate that massive postmerger ejecta associated with massive dynamical ejecta produce a smaller amount of Y as a result of low $Y_e$, which is realized by a lower neutrino flux in accordance with a short timescale for collapsing into BHs. We also note that there still exists a downward trend for [Y/Eu] with increasing [Eu/H], which might support that Y production is connected to an Eu production site instead of being correlated with Fe production as predicted by LEPP \citep{Travaglio_04, Montes_07}.

In our scheme, r-II stars are interpreted to be born from gas enriched by tidally-driven dynamical ejecta having very low $Y_e$, which result from either highly asymmetric double NS systems or BH-NS binaries. In such highly neutron-rich ejecta, the $Y_e$ distribution inside each ejecta is likely skewed to be so low that the mass fraction with very low $Y_e$ (e.g., $<0.18$) is, in some of them, high enough (e.g., $>30$\%) to induce a boost of actinide production \citep{Holmbeck_19}. This is consistent with the observed fact that among Th-measured halo stars, r-II stars include actinide-boost stars with higher fraction ($\sim$45\%) than that ($\sim$25\%) of less Eu-enhanced stars (r-I:+0.3$\leq$[Eu/Fe]$<$+1) \citep{Holmbeck_18}. These arguments point to the r-II star's origin which could demand a specific nucleosynthesis condition rather than the properties such as a mass-scale of protogalaxies (clouds) where these stars originally resided.

\section{Chemical evolution of r-process in the Galactic halo}

Chemical evolution study to explain an extensive scatter in [Eu/Fe] among Galactic halo stars utilizing cNSNSs has been implemented by many authors \citep[e.g.,][]{Argast_04, Voort_15, Shen_15, Hirai_15, Hirai_17}. The driver of a large scatter in these simulations is local inhomogeneities which can be produced when the limited amounts of interstellar matter are poluted by and mixed with the ejecta of each merger event occurring with a low frequency: inhomogeneous mixing could produce larger [Eu/Fe] in strongly polluted areas and smaller values in less polluted ones. On the other hand, 
our scheme predicts that the variation in [Eu/Fe] among halo stars is primarily caused by the different masses of tidally-driven dynamical ejecta from cNSNSs and cBHNSs. To validate this hypothesis,  we model the chemical evolution of the Galactic halo and calculate the Eu/Fe evolution.

We consider protogalactic fragments with a baryonic mass of $10^6$\msp, which is the minimum baryonic mass at the epoch of galaxy formation \citep{Tegmark_97}. This fragment mass is also implied from the threshold of an initial mass of protogalaxies, as implied from Ret II. The observed [Eu/H] $\approx-1$ of r-II stars in this galaxy implies that an Eu mass ejected from a cNSNS with $M_{\rm tid}\approx0.01$\ms is mixed with gas having a mass of $\sim 10^6$\ms \citep[see also][]{Ji_15, Ji_16}. In each fragment, separate chemical evolutions proceed as building blocks of the halo. Thus,  metal-poor stars in the Galaxy are predicted to be an assembly of stars originating from individual fragments, according to the hierarchical galaxy formation scheme. This scenario does not distinguish whether each fragment corresponds to an accreted unit of the Galaxy or one of many pieces of an accreted massive dwarf galaxy, such as Gaia-Enceladus \citep{Helmi_18}. In fragments, the ejecta are considered to propagate inside the whole volume and thus are  well-mixed. This is justified by both a broad uniformity of Eu abundances in Ret II \citep{Ji_16} and the abundance of $^{244}$Pu in the solar vicinity, which suggests that the ejecta of cNSNSs are mixed with local gas as massive as \gtsim$10^6$\ms \citep{Tsujimoto_17b}.

\begin{figure}[t]
	\includegraphics[width=\columnwidth]{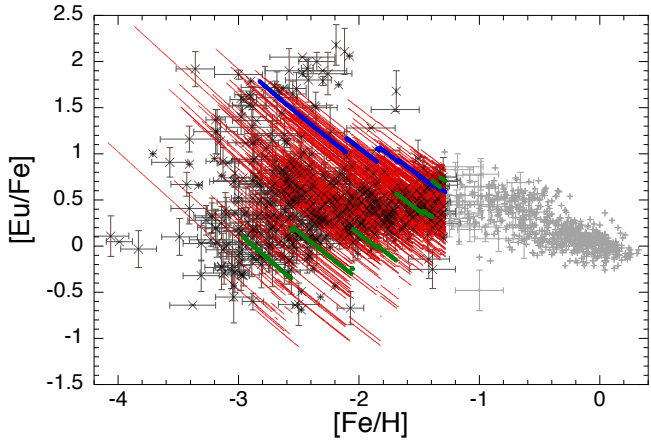}
\caption{Predicted [Eu/Fe]-[Fe/H] correlations for the Galactic halo (red lines) compared with the observed data (crosses). For reference, the data of mainly disk stars for [Fe/H]$>-1.3$ are shown by small pluses. Predicted feature is the assembly of 100 randomly selected cases for the chemical evolution in protogalactic fragments with a mass of $10^6$\msp. Among them, two representative evolutionary paths are highlighted by green and blue lines (see the text).}
\end{figure}

In each fragment, star formation is assumed to continue for a duration of 300 Myr at a low star formation rate so as to give $\langle$[Fe/H]$\rangle \approx -1.6$ as observed for halo stars, which results in a conversion of about 17\% of gas to stars. Owing to the short duration considered here, the contribution of Fe from type Ia SNe is not included. For Fe yields from CCSNe, we assume the nucleosynthesis yields tabulated in \citet{Kobayashi_06}. We assume that an Eu ejection occurs at a rate of one per $\sim$200 CCSNe (see \S 4.1), which leads to $\sim$6 events in total. For each event, the ejected mass of Eu is given by randomly assigning the mass of tidally-driven dynamical ejecta from $10^{-4}$\ms to 0.07 \msp, with a Gaussian probability distribution having a peak at $3\times10^{-3}$\ms for $M_{\rm tid}\leq 0.01$\msp. For $M_{\rm tid}\geq 0.01$\msp, we assign a 10\% probability in total with uniform mass distribution. Then, the Eu mass ejected from each merger event is calculated from $M_{\rm tid}$ composed of r-process elements with $Z\geq52$ following the solar r-process pattern. For the delay time of the ejection of r-process elements, we adopt a short timescale of $10-30$ Myr to accelerate the early r-process enrichment in the halo \citep{Tsujimoto_14b}. 

Since a sequence of r-process events occurs with randomly chosen masses of tidally-driven dynamical ejecta, individual evolutionary paths of Eu/Fe vary among protogalactic fragments. Figure 3 shows a mixture of such paths in 100 fragments. They include two examples that largely differ in Eu production by tidally-driven dynamical ejecta (distinguished by different colors): an initial Eu-boost making r-II stars, followed by a decreasing trend of [Eu/Fe] (blue), and a gradual increase in [Eu/Fe] via a sequence of mild Eu-enrichment by each event, resulting in no r-II stars (green). This figure excludes the initial generations of stars with no Eu content. It shows broad agreement between our predicted feature and the observation. On the other hand, if we adopt the delay time distribution for the merger objects spanning over gigayears implied from that for short gamma-ray bursts \citep{Fong_17}, there is a poorer agreement with the observation mainly due to a smaller predicted scatter in [Eu/Fe] by a reduced number of r-process events within a short duration (300 Myr) of star formation. Note that Eu abundance in disk stars favors a primary r-process production with a short time delay over long-delay merger events \citep[e.g.,][]{Cote_19, Siegel_19, Lin_19}. Accordingly, we claim that the large scatter in the abundance ratio that is unique to r-process elements in the Galactic halo is an end result of r-process enrichment driven by merger events under halo formation via mass assembly of small building blocks. In these merger events, widely spanning masses of tidally-driven dynamical ejecta are realized.

\section{Merger Event Rate}

Based on our proposed scheme, we can discuss the event rates of cNSNSs and cBHNSs associated with r-process production. On the other hand, Advanced LIGO/Virgo Observing Runs 1 and 2 by the ongoing gravitational-wave search project give the reliable rates of cNSNSs and cBHNSs (The LIGO Scientific Collaboration et al.~2018). Thus, the comparison between the two results will help to verify our argument presented here. In general, the assessment of merger event rates from the analysis of r-process abundance assuming the connection of r-process production with the merger events is highly uncertain unless a mean ejecta mass is narrowed down \citep{Rosswog_17, Hotokezaka_18}. Our scheme that assigns each (tidal) ejecta mass to each [Eu/Fe] ratio of Galactic halo stars could give a one possible way to resolve this problem through a statistical treatment.

\subsection{NS-NS}

First, we estimate the cNSNS event rate in the local volume (LV) of the Universe, $R_{\rm LV}$. This can be directly compared with the rate deduced from gravitational-wave events in the local Universe. $R_{\rm LV}$ can be converted from the present rate for the Galaxy, $R_{\rm G}$, which is obtained through an analysis of stellar r-process abundances combined with the recent star formation activity in the Galaxy \citep{Tsujimoto_14b, Tsujimoto_17b}. Note that $R_{\rm G}$ would be lower than the mean rate for the Galaxy \citep{Rosswog_17, Hotokezaka_18} owing to the currently low star formation rate \citep[e.g.,][]{Mor_19}, and this tendency of star formation seen in the Galaxy is shared with M31 \citep{Bernard_15}. In this study, we utilize the most reliably measured stellar abundance: the solar abundance.

The key point is that the relative solar abundance ratio of Eu to Mg ($5.2\times 10^{-7}$) is equivalent to the relative nucleosynthesis yield between the two elements. Here the Eu yield is defined per CCSN basis, that is, as the combination of the Eu mass per cNSNS event and a reciprocal number of CCSNe that is necessary to give birth to one cNSNS event. For the Mg yield from a single CCSN, we adopt a mass of 0.1 \msp. This value is deduced from two arguments: (i) nucleosynthesis calculations for CCSNe give 0.1$-$0.125 \ms as an IMF-weighted average Mg mass for the progenitor's mass range of 13$-$50 \ms \citep{Tominaga_07} and (ii) a relative yield with a mean Fe mass of $\sim$0.07 \ms obtained from a light curve analysis of CCSNe \citep{Hamuy_03} leads to an observed plateau of [Mg/Fe]$\approx$+0.4 among halo stars and thus the value of the most metal-poor disk stars \citep[e.g.,][]{Bensby_14}. Accordingly, we obtain the Eu yield of $5.2\times 10^{-8}$\msp.

Then, this together with an average Eu mass per cNSNS event leads to the cNSNS rate. The Eu mass fraction of $6\times10^{-3}$ in the tidally-driven dynamical ejecta, which is obtained from the assumption of the solar r-process pattern for the elements with $Z\geq$52, provides the cNSNS rate per CCSNe of one per 1150$(\langle M_{\rm tid}\rangle$/0.01\msp) CCSNe. This further leads to the Galactic cNSNS rate $R_{\rm G}$, using the present-day Galactic CCSN rate of 2.3 SNe per century \citep{Li_11}:
\begin{equation}
R_{\rm G} \approx 20 \left(\frac{\langle M_{\rm tid}\rangle}{0.01 M_\odot}\right)^{-1} \ \ {\rm Myr}^{-1}.
\end{equation}

\noindent Finally, this rate can be converted to the rate for the local volume of the Universe via the density of a Milky Way-equivalent galaxy of $1.16\times 10^{-2}$ Mpc$^{-3}$ estimated from the blue luminoisty \citep{Abadie_10}:

\begin{equation}
R_{\rm LV} \approx 230 \left(\frac{\langle M_{\rm tid}\rangle}{0.01 M_\odot}\right)^{-1}\ \ {\rm Gpc}^{-3}{\rm yr}^{-1}.
\end{equation}

\noindent This $R_{\rm LV}$ range is within the rate at the 90\% confidence intervals of $110-3840$ Gpc$^{-3}$ yr$^{-1}$,  with the most probable value around $\sim 1000$~Gpc$^{-3}$ yr$^{-1}$ by gravitational-wave detection \citep{LIGO_18}. Therefore, $R_{\rm LV}$ deduced from the solar abundance is in agreement with that from the gravitational-wave event rate if the average mass of tidally-driven dynamical ejecta is assumed to be a few times $10^{-3}$ \msp.

\subsection{BH-NS producing r-process elements}

The cBHNSs events associated with r-process production are considered rare among all cBHNSs \citep{Zappa_19} whose event rate is constrained to be $<610$ Gpc$^{-3}$ yr$^{-1}$ \citep{LIGO_18}. We can predict the approximate rate based on our theoretical scheme that the group of stars enriched by the r-process to the highest level ([Eu/Fe]\gtsim$+1.4$) can be connected to a cBHNS origin, while other stars are associated with a cNSNS origin. The fraction of r-II stars compared to all metal-poor halo stars is estimated to be about $2-4$\%, although this estimate likely overrepresents due to incomplete sampling \citep{Brauer_19}. Then, considering that the fraction of r-II stars with [Eu/Fe]$\geq +1.4$ with respect to all r-II stars is about $25$\% \citep[10/38:][]{Roederer_18}, the event ratio of cBHNSs producing r-process elements to cNSNSs is implied to be about $0.5-1$\% at maximum. Accordingly, we anticipate that the future detection of the gravitational-wave events originating from cBHNSs could be inclined towards no association with electromagnetic counterparts. If we adopt $1000$~Gpc$^{-3}$ yr$^{-1}$ as a fiducial value of the cNSNS event rate, the event rate of cBHNS with the r-process, a few per Gpc$^{3}$ per year, is deduced as the most likely case.

\section{Summary}

We show that the large scatter in the [Eu/Fe] ratio among Galactic halo stars can be the outcome of varying Eu production,  spanning three orders of magnitude in the masses of tidally-driven dynamical ejecta from both cNSNSs and cBHNSs. The origin of r-II stars is predicted to have a close connection to a highly neutron-rich environment, which could be realized in the massive ($\sim0.01-0.02$ \msp) dynamical ejecta from cNSNSs with highly asymmetric double NS systems or those from cBHNSs. On the other hand, enrichment by cNSNSs with less massive ejecta results in halo stars having an observed variation in the [light r-process/Eu] ratio with less enhancement in Eu. We further predict that the typical mass of tidally-driven dynamical ejecta for cNSNSs should be a few times $10^{-3}$ \msp, which is nicely supported by the cNSNS rate deduced from the ongoing gravitational-wave observations.

Our scenario is based on the idea that the correlation between Y/Eu and Eu/Fe emerges from different neutron richness of the ejecta caused by different properties of coalescing binaries. Specifically, the amount of the neutron-rich tidal ejecta and the Eu yield are presumed to increase as the binary becomes asymmetric and tidal deformation of the lighter component is enhanced. Although this trend is consistent with the results of previous studies \citep[e.g.,][]{Sekiguchi_16}, whether the observed correlation is reproduced quantitatively should be investigated by future hydrodynamical simulations and nucleosynthetic calculations with systematically varying the binary parameters such as the mass ratio. These inputs would help to promote a thorough understanding of Galactic elemental features for all r-process elements, including light r-process nuclei, lanthanides, and actinides.

\acknowledgements

The authors thank an anonymous referee for his or her valuable comments. This work was supported by JSPS KAKENHI grant Nos.~16H06342, 17H01131, 18H01258, 18H04593,  18H04595, JP18H05236, and 19K14720, and MEXT Japan ("Priority Issue on Post-K computer: Elucidation of the Fundamental Laws and Evolution of the Universe"). NN thanks S. Fujibayashi for fruitful discussions on the post-merger ejecta.

\end{document}